\documentclass{article}

\usepackage{PRIMEarxiv}

\usepackage[utf8]{inputenc} 
\usepackage[T1]{fontenc}    
\usepackage{hyperref}       
\usepackage{url}            
\usepackage{booktabs}       
\usepackage{amsfonts}       
\usepackage{nicefrac}       
\usepackage{microtype}      
\usepackage{lipsum}
\usepackage{fancyhdr}       
\usepackage{graphicx}       
\graphicspath{{media/}}     
\usepackage{amsmath}
\usepackage{amssymb}
\usepackage{graphicx}
\usepackage{cite}
\usepackage{amsfonts}\usepackage{algorithmic}
\usepackage{textcomp}
\usepackage{xcolor}
\usepackage{subfig}

\title{EdgePrompt: A Distributed Key-Value Inference Framework for LLMs in
6G Networks
\thanks{Jiahong Ning, Tingting Yang and Pengyan Zhu are with Dalian Marine University, School of navigation, China (Email: yangtingting820523@163.com, jiahong.ning@mnsu.edu). Ce ZHENG is with Pengcheng Lab, Shenzhen, China (zhengc@pcl.ac.cn). Sumei Sun, and Gary Lee are with the Institute for Infocomm Research (I2R), A{*}STAR, Singapore (e-mail: sunsm@i2r.a-star.edu.sg, gary\_lee@i2r.a-star.edu.sg), Jiahong Ning is also with the Institute for Infocomm Research (I2R), A{*}STAR, Singapore (stuningj@i2r.a-star.edu.sg), and Pengcheng Lab, Shenzhen, China (ningjh@pcl.ac.cn). This work has been accepted by the IEEE INFOCOM 2025 Workshop on PerAI-6G. The final publication will be available via IEEE Xplore} 
}

\author{
 Jiahong Ning, Pengyan Zhu, Ce Zheng, Gary Lee, Sumei Sun, Tingting Yang
}

\begin{document}
\maketitle

\begin{abstract}
As sixth-generation (6G) networks advance, large language models (LLMs) are increasingly integrated into 6G infrastructure to enhance network management and intelligence. However, traditional LLMs architecture struggle to meet the stringent latency and security requirements of 6G, especially as the increasing in sequence length leads to greater task complexity. This paper proposes EdgePrompt, a cloud-edge collaborative framework based on a hierarchical attention splicing mechanism. EdgePrompt employs distributed key-value (KV) pair optimization techniques to accelerate inference and adapt to network conditions. Additionally, to reduce the risk of data leakage, EdgePrompt incorporates a privacy preserving strategy by isolating sensitive information during processing. Experiments on public dataset show that EdgePrompt effectively improves the inference throughput and reduces the latency, which provides a reliable solution for LLMs deployment in 6G environments.
\end{abstract}


\section{Introduction}
The advent of Large Language Models (LLMs), such as GPT-4 and LLaMA, has revolutionized the field of natural language processing, excelling in tasks such as text generation, question answering, and conversational systems \cite{1}. These models have been widely adopted across various industries, including education, healthcare, and finance. As we approach the era of sixth-generation (6G) communication networks, the integration of LLMs into 6G infrastructure presents a transformative opportunity. Integrating LLMs within 6G architecture could enable intelligent network management, data analytics, and personalized experiences. However, deploying LLMs in the 6G infrastructure also introduces several challenges. 

First, LLMs inference incurs significant latency due to computational demands introduced by sequential processing \cite{28,58_network-LLM}. This latency poses a critical challenge for real-time 6G applications such as autonomous driving and virtual reality. Second, transmitting user data to cloud-based LLMs for inference raises severe privacy issues, because sensitive information may be exposed during transmission and processing \cite{42_leveraging_ai_review,41}.

Recently, researchers focus on model compression technology and the deployment on the edge, in order to reduce network latency and enhance data security. According to \cite{12-llmservey}, LLMs inference is often hindered by high memory access demands or insufficient compute resources, referred to as being memory-bound or compute-bound, respectively. 
To tackle the memory-bound issue, \cite{llm_qat} proposed the quantization aware training (QAT), which embeds low-precision representations directly into the training pipeline, providing faster device inference with minimal accuracy loss. 
The paper \cite{lut_gemm} leverages weight-only quantization to optimize matrix multiplications in LLMs, achieving reduced latency and enhanced computational efficiency. However, in the transformer architecture, both the quadratic complexity of self-attention with sequence length and the increased computational load of large batches shift the bottleneck from memory bound to compute bound. Even aggressive quantization fails to alleviate system limitations.

Quantization reduces model size and memory usage to accelerate inference in memory-bound scenarios \cite{63_roofline_model}, but offers limited benefits in compute-bound cases. As a result, researchers resort to parallel decoding approaches. One of the viable solutions is speculative sampling \cite{5-SS}, where a sequence of candidate tokens is generated serially by the draft model and then verified in parallel by the target model. This significantly accelerate inference without any change to the model architectures, training regimes and output distributions. 

Accordingly, a new paradigm of cloud-edge collaboration was proposed: a fast small model with low computational cost resides on device, while a large language model is deployed in the cloud, which has sufficient memory and computational resources\cite{65-edge,66_hybrid,67_EdgeLLM}. 
However, the inference acceleration achieved through speculative decoding should offset the latency introduced by data exchange between cloud and edge. What is more, cloud-edge interactions increase the risk of sensitive data exposure.

It introduces additional latency and synchronization challenges due to frequent data exchanges and alignment requirements. And the cloud-edge interactions increase the risk of sensitive data exposed, which limited the envision of 6G network. To address this problem, we propose \textbf{EdgePrompt}, a novel cloud-edge collaborative inference architecture for communication networks.

To balance inference efficiency and privacy protection, we propose \textbf{EdgePrompt}, a novel cloud-edge collaborative inference architecture for communication network. Unlike traditional task offloading, EdgePrompt separates sensitive data from cloud prompts, ensuring that sensitive information stays on edge devices, while leveraging powerful cloud
resources for large-scale computation. EdgePrompt not only enhances privacy protection but also improve bandwidth and compute bottlenecks. By the high throughput and low latency of 6G networks, EdgePrompt  dynamically adapts to varying traffic and workload patterns. Moreover, as a complementary solution, EdgePrompt complements existing compression and parallelization strategies.

\begin{figure}
\vspace{-0.2cm}
\subfloat[The workflow of EdgePrompt \label{fig:The-workflow-of}]{\includegraphics[scale=0.5]{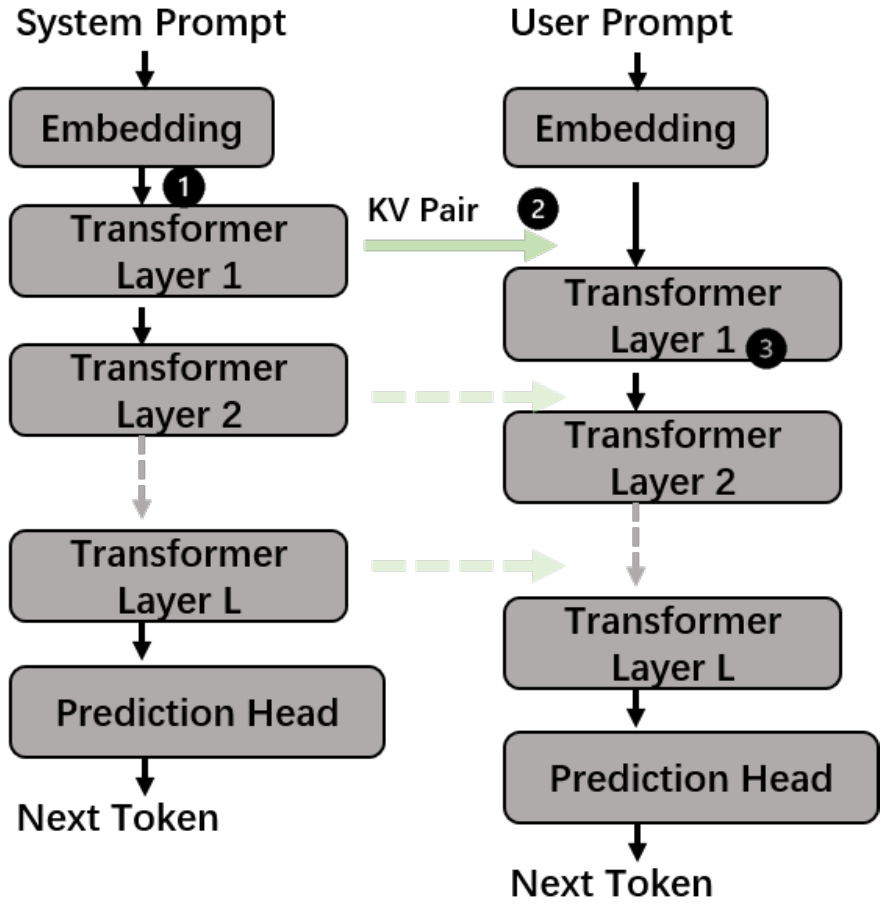}}

~~~~~~~\subfloat[Attention structure of EdgePrompt \label{fig:The-attention-structure}]{\includegraphics[scale=0.7]{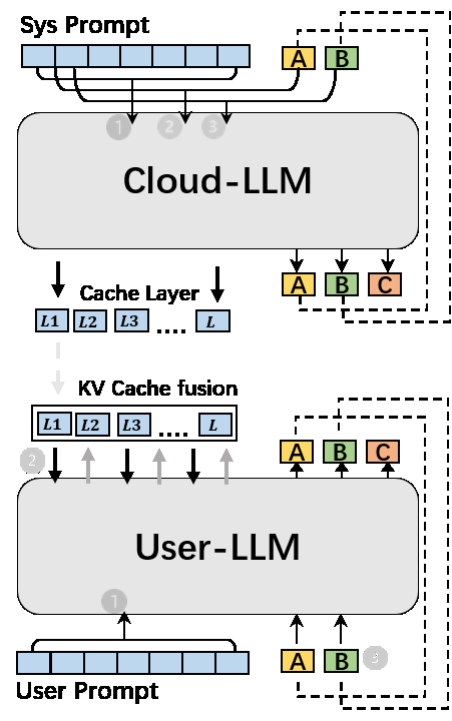}}

\caption{The EdgePrompt architecture}
\label{fig:model desgin}

\vspace{-0.2cm}
\end{figure}

As shown in Fig.\ref{fig:The-workflow-of}, the EdgePrompt architecture
ensures that sensitive data remain on local devices, significantly
mitigating privacy risks by localized data processing. By partitioning
the computationally intensive processing of cloud prompts to the cloud, EdgePrompt balances the computational load on edge devices.
Additionally, the exchange of KV pair minimizes communication overhead,
resulting in decreased inference latency and improved throughput compared
to traditional scheme. This strategic utilization facilitates the
efficient handling of complex LLM tasks in 6G networks, thereby meeting
the demands of applications. 

Our primary contributions are as follows:
\begin{itemize}
\item \textbf{Proposed a novel architecture for inference efficiency and
privacy protection:} We introduce EdgePrompt, a cloud-edge collaborative
inference framework designed for LLMs foundation model into communication
networks. By separating cloud and edge prompt and managing KV
Cache locally, this design enhances inference efficiency while ensuring
privacy.
\item \textbf{Theoretical and empirical analysis:} We conduct a detailed
theoretical and performance analysis of EdgePrompt, including the
establishment of optimization models and rigorous modeling specific
to communication network environments. Our analysis demonstrates the
practicality and effectiveness of EdgePrompt in optimizing LLMs inference
under cloud-edge collaboration.
\item \textbf{Performance improvements:} We design experiments and analysis
to demonstrate that EdgePrompt achieves significant improvements in
system throughput and user data privacy within 6G networks, providing
a robust solution for deploying LLMs in privacy sensitive and latency
critical applications.
\end{itemize}

\section{Methodology\label{sec:Methdology}}

In this section, we elaborate on the models and methods we have established.
Section \ref{subsec:Analysis-of-the} provides a brief analysis of
the inference process. Section \ref{subsec:Idea-description} delves
into the mathematical expressions to analyze the problems and presents
our solutions. Section \ref{subsec:Communication-Model} details the
design and analysis of the EdgePrompt system modules with time slot. 

\subsection{The processing of inference\label{subsec:Analysis-of-the}}

The inference process of LLMs follows an autoregressive generation
paradigm, where each token is generated sequentially based on preceding
tokens in the context. This process comprises two stages: \textbf{Prompting stage}, where input sequences are processed in parallel to generate
query $(\ensuremath{Q})$, key $(\ensuremath{K})$, and value $(\ensuremath{V})$
vectors efficiently. \textbf{Autoregressive generation stage}, where
tokens are generated step by step, limits parallelization and dominates the computational load. As sequence lengths grow, the demand on memory and bandwidth for the KV also increases, resulting in significant latency, particularly for concurrent tasks.

To address the computational inefficiency of long sequences, techniques like PagedAttention and advanced cache management have been introduced \cite{6,7-MLsys}. PagedAttention divides sequences into smaller pages, computing attention independently within each page.  This approach reduces memory usage and lowers computational complexity from quadratic to linear. For attention computation, the self-attention mechanism is:
\begin{equation}
\mathbf{A}=\text{softmax}\left(\frac{\mathbf{Q}\mathbf{K}^{T}}{\sqrt{d}}\right),\quad\mathbf{O}=\mathbf{A}\mathbf{V},
\end{equation}
where $Q$, $K$, and $V$ represent token dependencies, and $d$ is the dimensionality. While PagedAttention enhances efficiency, it may cause context fragmentation, degrading the performance of inference tasks due to lack of global information.

Cache management extends these optimizations by precomputing and storing modular attention states for recurring prompt segments. These precomputed states are represented:
\begin{equation}
\mathbf{A}_{\text{module}}\!\!=\!\!\text{softmax}\!\left(\!\!\frac{\mathbf{Q}_{\text{module}}\mathbf{K}_{\text{module}}^{T}}{\sqrt{d}}\!\!\right), ~ \mathbf{O}_{\text{module}}\!\!=\!\!\mathbf{A}_{\text{module}}\!\mathbf{V}_{\text{module}}.
\end{equation}

Attentions are reused across inference steps, reducing redundant computations and enhancing efficiency for multi-step tasks. This approach significantly improves throughput, particularly in tasks involving frequent reuse of similar prompts, while keeping the coherence and quality of generated outputs.

\subsection{The EdgePrompt
\label{subsec:Idea-description}}

Researchers have made efforts to address the memory-bound issues caused
by I/O operations. In [10], the authors proposed solutions for GPU RAM management, aimed at mitigating memory wastage during inference and reducing I/O operations by reusing caches.
Building on these foundational works, our research further explores how edge devices enhance computational efficiency and reduce inference latency while preserving user privacy. 

As illustrated in  Fig.\ref{fig:model desgin}, we consider a cloud-edge collaborative scenario. In this setup, a cloud server with ample computational resources supports edge devices in handling inference tasks. LLMs inference generally includes two phases: (1) the prompt phase, where hidden states for the entire prompt (both cloud and edge) are computed to generate the first token; (2) the auto-regressive generation phase, where each subsequent token is generated sequentially. EdgePrompt optimizes the auto-regressive generation phase that constitutes the main response process.

The core concept of EdgePrompt lies in separating the cloud prompt from the edge prompt, with the cloud handling cloud prompt computations and the edge processing edge prompts. This approach synchronizes KV pair intermediate vectors to maintain computational consistency.
As depicted in Fig.\ref{fig:The-attention-structure}, an entire prompt consists of cloud prompts and edge prompts. The cloud prompt activates the LLMs capabilities, while the edge prompt provides specific, individual request information. Ensuring high-quality responses
requires a detailed and professional cloud prompt, which is known
as prompt engineering. Typically, the cloud prompt should be comprehensive
and intricate to steer the model's output according to predefined
logic. The edge prompt, containing substantial personal information,
is integral to privacy protection.

The LLMs model typically features a multi-layered architecture, which includes an embedding layer like Fig.\ref{fig:The-workflow-of}, multiple decoder layers, an output layer, and various other modifications.
During inference, KV pairs are generated, with each layer's
output serving as the input for the subsequent layer. 
Fig.\ref{fig:model desgin} shows how EdgePrompt distributes cloud
prompt and edge prompts. 

In this multi-layer architecture, each layer $l$ operates using query,
key, and value vectors, denoted as $\mathbf{Q}^{l}$, $\mathbf{K}^{l}$
and $\mathbf{V}^{l}$. The query vector $\mathbf{Q}^{l}$ represents
the model\textquoteright s current focus for layer $l$, the key vector
$\mathbf{K}^{l}$ captures the relevance of previous information,
and the value vector $\mathbf{V}^{l}$ contains the actual data to
be aggregated based on calculated attention scores. This computation
occurs separately for the cloud and edge prompts, ensuring that each
prompt type can influence the model\textquoteright s output in independent ways. The attention matrix of cloud prompt in each layer is given as:
\begin{equation}
\mathbf{A^{\mathit{l}}}_{cloud}=softmax\left(\frac{\mathbf{Q}_{cloud}^{l}\mathbf{K^{\mathit{l}}}_{cloud}^{T}}{\sqrt{d}}\right)\mathbf{V}_{cloud}^{l},\label{eq:system attention}
\end{equation}
and the attention matrix of edge prompt in each layer is:
\begin{equation}
\mathbf{A}_{edge}^{\mathit{l}}=softmax\left(\frac{\mathbf{Q^{\mathit{l}}}_{edge}\mathbf{K^{\mathit{l}}}_{edge}^{T}}{\sqrt{d}}\right)\mathbf{V^{\mathit{l}}}_{edge}.\label{eq:edge attention}
\end{equation}

In the fusion process, each layer $l$ generates an output $\mathbf{O}_{module}^{l}$ by combining the attention matrices from the cloud and edge prompts, leveraging KV pair for efficient synchronization. Specifically, the cloud server computes the attention $\mathbf{A^{\mathit{l}}}_{cloud}$ for the cloud prompts. The KV pair is cached at the cloud server and sent to the edge device in the meantime.
This approach ensures that essential information is efficiently shared
without redundant computation, as the edge device receives $\mathbf{A^{\mathit{l}}}_{cloud}$
and combines it with the locally computed attention matrix $\mathbf{A^{\mathit{l}}}_{edge}$
in layer $l$. The weights $\alpha_{cloud}$ and $\alpha_{edge}$
controlling the influence of prompts type, determine the respective contributions
of the cloud and edge prompts at each layer, and we have the fusion formulate:
\begin{equation}
\mathbf{O}_{module}^{l}=\alpha_{cloud}\mathbf{A}_{cloud}^{l}+\alpha_{edge}\mathbf{A}_{edge}^{l}.\label{eq:fusion processing}
\end{equation}

The KV pair thus enables synchronized and efficient fusion, to produce
$\mathbf{O}_{module}^{l}$. By leveraging the cloud for computationally
intensive cloud prompts while managing user-specific data locally,
EdgePrompt optimizes both performance and privacy. By offloading system
computations to the cloud and handling sensitive data locally, EdgePrompt
achieves computational efficiency while preserving user privacy. 

\subsection{Communication Model \label{subsec:Communication-Model}}

\begin{figure}
\vspace{-0.1cm}
\includegraphics[scale=0.45]{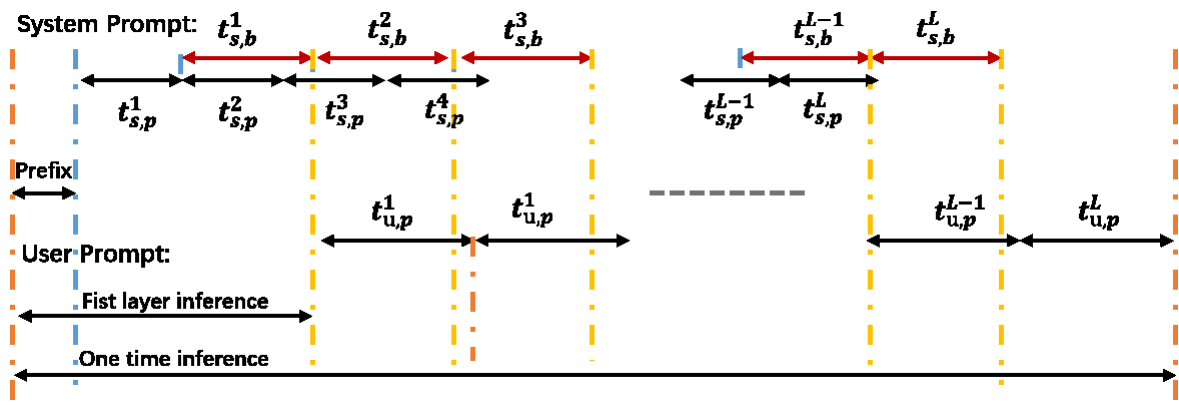}

\caption{A complete process of Edge prompt}
\label{fig:A-complete-process}

\vspace{-0.3cm}
\end{figure}

In Section \ref{subsec:Communication-Model}, we introduce the architecture
of EdgePrompt. Here, we analyze EdgePrompt from a time-scale perspective
to optimize cloud-edge computation and communication, aiming to minimize
the overall inference delay $T$. To achieve this, we employ an overlapping
timeline, enabling concurrent data transmission and computation. This
overlapping approach allows the system to perform data transmission
and computation in parallel, thereby reducing waiting times and maximizing
resource utilization. 


As shown in the Fig.\ref{fig:A-complete-process}, we break down the inference timeline into several key components. The initial upload request is minimal, which is included in the parameter initialization
time, denoted as $T_{prefix}$, representing the setup phase before inference begins. $t_{c,c}^{l}$ and $t_{e,c}^{l}$ represent the computation time of the cloud and the edge device, respectively. $t_{c,t}^{l}$ represents the time required to transmit the KV pair from cloud to edge.


The total inference time for the cloud prompt and edge prompt, denoted as $T_{cloud}$ and $T_{edge}$ respectively, are
\begin{equation}
\begin{cases}
T_{cloud}=(t_{c,c}^{1}+t_{c,t}^{1})+\sum_{l=2}^{L}\left(t_{c,c}^{l}+t_{c,t}^{l}\right)\\
T_{edge}=(t_{c,c}^{1}+t_{c,t}^{1})+\sum_{l=1}^{L} t_{e,c}^{l}\end{cases},
\end{equation}
where $T_{cloud}$ are composed of the computation and communication time of all layers at cloud, and $T_{edge}$ should include the cloud compuation and transmission time at the first layer.

To achieve the minimum overall inference time $T$, we consider the initial setup time $T_{prefix}$, the first layer\textquoteright s computation and transmission time, and an optimal overlap of computation and transmission across remaining layers. The formulated is:
\begin{align}
\min T=& T_{prefix}+(t_{c,c}^{1}+t_{c,t}^{1})+\nonumber \\
&\min\{\max[\sum_{l=2}^{L}\left(t_{c,c}^{l}+t_{c,t}^{l}\right),\sum_{l=1}^{L-1}\left(t_{e,c}^{l}\right)]+t_{e,c}^{L}\} & .
\end{align}

We adopt the overlapping approach, which simultaneously handles computation performance and communication behavior. Through parallel processing, system performance is optimized, significantly reducing time consumption. The validity of this model is verified and implemented based on the
following assumptions and theoretical frameworks.

\textbf{Assumption 1}: The computational speed of the cloud server exceeds that of the edge device, expressed as $F_{sys}\geq F_{edge}$, where $F_{sys}$ and $F_{edge}$ denote their respective FLOPS of GPU frequencies.

This implies that the cloud server is faster than the edge device under the same workload, i.e. $t_{c,c}^{l} < t_{e,c}^{l} $.


\textbf{Assumption 2}: The workload $C^{l}$ remains
approximately the same for each layer between cloud server and edge devices. 

In the transformer architecture, the amount of computation per layer is mainly determined by the model parameters such as weight matrix and hidden layer dimensions. Therefore, without huge gap in prompt lengths, the workload $C^{l}$ remains approximately the same for each layer between cloud and server devices.

Additionally, each decoding layer is assumed to maintain
a consistent KV pair size, calculated as: $KV~\mathrm{size} = 2\times H\times D\times S$,
where $H$ is the number of attention heads, $D$ is the hidden
layer dimension, and $S$ is the sequence length. This consistency
in cache size implies stable KV pair transmission time $t_{c,t}^{l}$,
and edge prompt computation time $t_{e,c}^{l}$ across layers:
\begin{align}
t_{c,t}^{l}\approx t_{c,t}^{l-1} ,\\
t_{e,c}^{l}\approx t_{e,c}^{l-1}.
\end{align}

According to Assumptions 1 and 2, we have the following two conclusions:

\textbf{Conclusion 1:}
The computation time
for the cloud prompt\textquoteright s KV pair at any layer $l$
on the cloud server is shorter than the computation time for the edge
prompt\textquoteright s KV pair on the same layer at the edge. Formally,
this relationship is expressed as:
\begin{align}
    t_{c,c}^{l}<t_{e,c}^{l}.
\end{align}

With the overlaping approach, we identify two divergent conditions: the computation time for the cloud $t_{c,c}^{l}$ and the communication time $t_{c,t}^{l}$. When $t_{e,c}^{l}<t_{c,t}^{l}$, under the \textbf{Conclusion 1} and \textbf{Assumption 2}, we derive the following relationship:
\begin{align}
    t_{c,c}^{l}<t_{e,c}^{l-1}<t_{c,t}^{l-1} .   
\end{align}

Consequently, we can model the current conditions. As shown in Fig.\ref{fig:A-complete-process},
the primary focus is on the auto-regressive generation process, where
the main factor determining the overall inference time $T_{inf}$
is the cumulative communication time: $T_{inf}=\sum_{l=1}^{L}\left(t_{c,c}^{l}\right)$.
By optimizing the user's channel condition, we can significantly reduce
$T_{inf}$. Thus, we have:
\begin{align}
P1: & \underset{t_{c,t}^{l}}{\min}\left(T_{prefix}+(t_{c,c}^{1}+t_{c,t}^{1})+\sum_{l=2}^{L}\left(t_{c,t}^{l}\right)+t_{e,c}^{L}\right).\label{eq:P1}
\end{align}

In (\ref{eq:P1}), the objective is to minimize the total inference
time $T_{inf}$ when communication time $t_{c,t}^{l}$ dominates
the computation time. Here, $T_{prefix}$ represents the initial setup time. The term $t_{s,p}^{1}+t_{s,b}^{1}$
accounts for the computation and transmission times in the first layer,
where both operations contribute substantially to the initial processing.
Starting from $l=2$, we focus on the cumulative communication time
across layers, $\sum_{l=2}^{L}\left(t_{c,t}^{l}\right)$, which becomes
the primary bottleneck as the model proceeds through additional layers.
Finally, $t_{e,c}^{L}$ represents the edge prompt computation time
at the final layer, marking the completion of the inference process.

Similarly, while $t_{c,c}^{l}<t_{c,t}^{l-1}<t_{e,c}^{l-1}$, the main
factor determining $T_{inf}$ shifts to the cumulative computation
time for the edge prompt: $T_{inf}=\sum_{l=1}^{L}\left(t_{u,p}^{l}\right)$.
In this case, by optimizing transmission time, the inference
processing can be significantly accelerated. Therefore, the optimization
goal becomes:
\begin{align}
P2:\underset{t_{e,c}^{l}}{\min} & \left\{ T_{prefix}+(t_{c,c}^{1}+t_{c,t}^{1})+\sum_{l=1}^{L}\left(t_{e,c}^{l}\right)\right\} .\label{eq:p2}
\end{align}

Similar to (\ref{eq:P1}), the expression includes $T_{inf}$ and
$t_{c,c}^{1}+t_{c,t}^{1}$, which together account for the initial
setup and processing of the first layer. However, the summation $\sum_{l=1}^{L}\left(t_{u,p}^{l}\right)$ starts
from $l=1$, because all layers contribute to edge prompt processing.
Since edge prompt computation becomes the limiting factor in (\ref{eq:p2}),
optimizing edge device performance to expedite edge prompt processing
in all layers can reduce the overall inference time.

\textbf{Conclusion 2}: Given that $t_{c,t}^{l}\approx t_{c,t}^{l-1}$, $t_{c,c}^{l}<t_{e,c}^{l}$ from \textbf{Assumption 1} and \textbf{Assumption 2}, the condition $t_{c,t}^{l}+t_{c,c}^{l}<t_{c,t}^{l-1}+t_{e,c}^{l}$ hold.

When the cloud-edge communication condition is better than calculation, then $t_{c,t}^{l-1}<t_{c,c}^{l}<t_{e,c}^{l-1}$, we can have the time condition \textbf{Conclusion 2}, then model could be present as:
\begin{align}
P3:\underset{t_{e,c}^{l}}{\min} & \left\{ T+(t_{c,c}^{1}+t_{c,t}^{1})+\sum_{l=1}^{L}\left(t_{e,c}^{l}\right)\right\} .
\end{align}

According to the above analysis, we can optimize the efficiency of hierarchical transmission according to the network condition and task complexity.

\section{Experiments\label{sec:Experiments}}

In this section, we conduct experiments to evaluate what the effect of EdgePrompt architecture on inference performance. We provide the experimental setup in Section \ref{subsec:Experimental-Setup}. Our
major evaluation is two scenarios: non-interactive batch processing
in Section \ref{subsec:Noninteractive-Batch-Processing} and interactive
service Section \ref{subsec:Interactive-Serving}. We use the \textbf{Llama-2-7b}
model for evaluation throughput and delay. 

\subsection{Experimental Setup\label{subsec:Experimental-Setup}}

This experiment aims to evaluate the performance of the distributed inference framework EdgePrompt in terms of inference throughput and latency, and to compare with the existing acceleration techniques. We chose to base our experiments on the ShareGPTv3 dataset, which contains 53,000 real conversations between users and ChatGPT, covering a wide range of topics and context lengths, and the Llama2-7b model. It is an ideal benchmark to evaluate reasoning performance on long and short cue tasks. The experimental environment includes two GPU platforms: NVIDIA GTX 4090 and NVIDIA A800. The GTX 4090 is a high-performance GPU widely used in edge computing scenarios, while the A800 is a data center GPU suitable for high-performance computing and AI inference tasks. The two kinds of hardware can reflect the powerful performance of cloud computing, which provides support for comprehensive testing of different inference methods.

In this section, we contrast three inference schemes, namely PageAttention, RelayAttention, and EdgePrompt. PageAttention is an open source LLMs inference acceleration technology. The core of PageAttention is to manage the memory requirements of large model inference by memory
paging mechanism, and optimize GPU memory utilization by efficient scheduler, so as to achieve high throughput service. RelayAttention is further optimized on the basis of PageAttention, and makes full use of the kv pair prompted by the system, which not only significantly improves the inference throughput, but also better deals with the memory bottleneck problem in long sequence tasks. EdgePrompt, as a distributed inference framework proposed by us, offloads the computing task of cloud prompts to the cloud through the cloud-edge collaboration mechanism, and processes edge prompts on the edge device at the same time, so as to effectively reduce the inference delay and enhance the privacy protection ability while improving the system throughput.

\begin{figure}
\vspace{-0.1cm}
\centering
\subfloat[Servering on the Cloud]
{
\includegraphics[scale=0.2]
{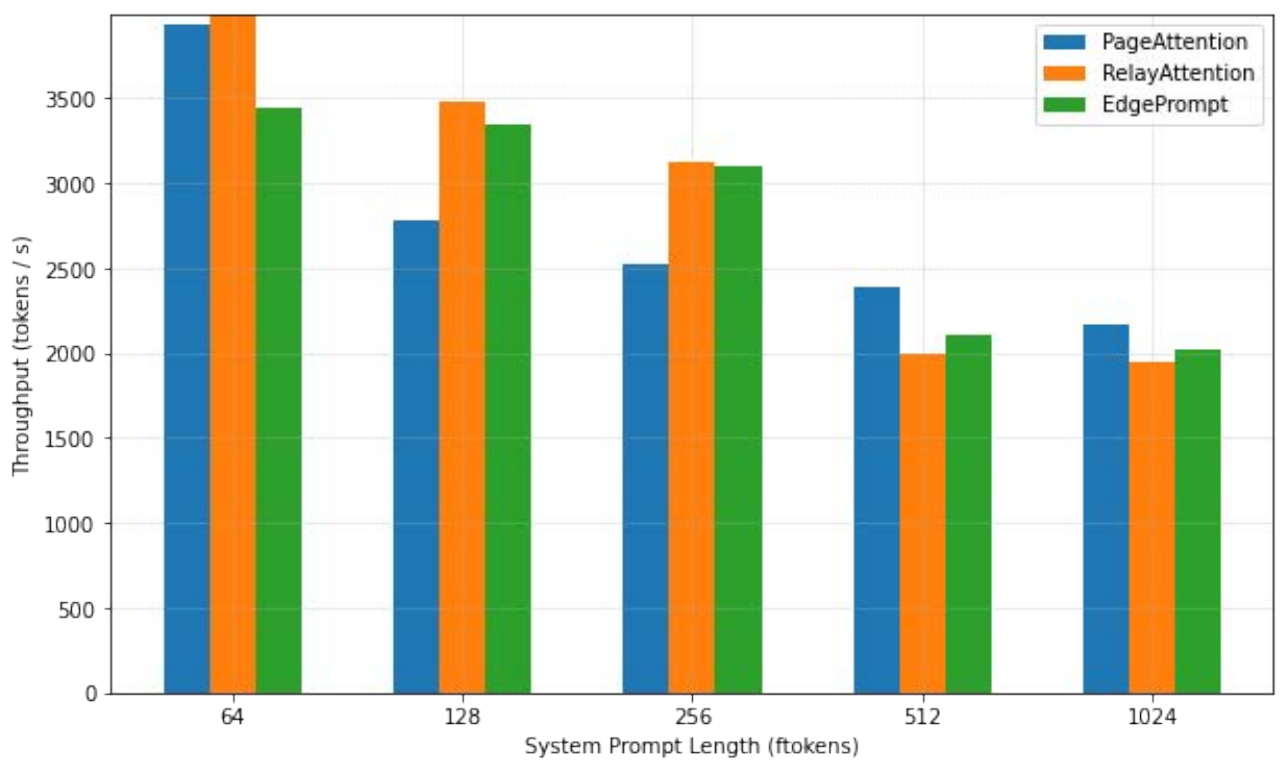}
}
\subfloat[Servering on the Edge]
{
\includegraphics[scale=0.2]
{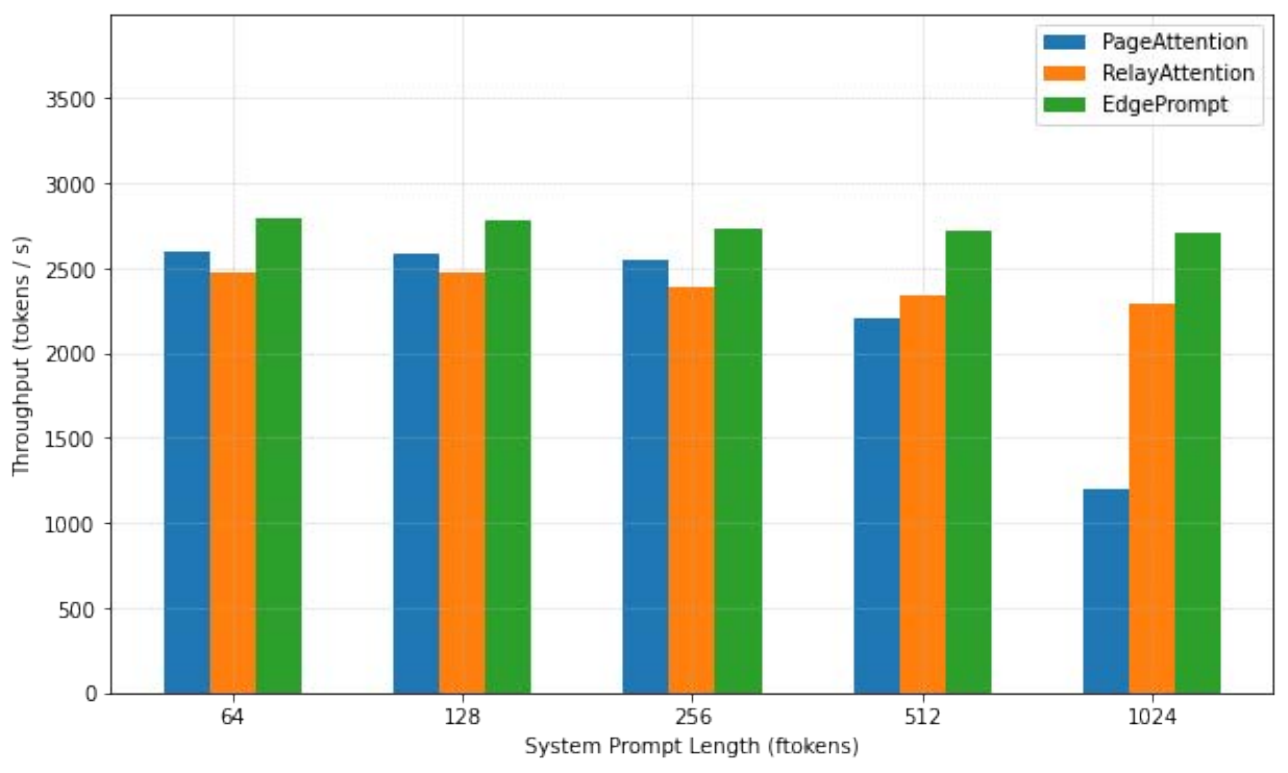}
}

\caption{The throughput in non-interactive processing}
\label{fig:The-GPU-throughput}

\vspace{-0.1cm}
\end{figure}

\subsection{Non-interactive Batch Processing\label{subsec:Noninteractive-Batch-Processing}}

For non-interactive batching, the user submits a request and retrieves
the results. PageAttention and RelayAttention are deployed in the
cloud or the edge, while our scheme adopts the cloud-edge collaborative
inference method. We measure throughput ($Token/s$) and processing
time. In a grouping manner, we divided 1000 requests into 100 groups
and submitted them to the model in turn to simulate the performance
loss of the model when filling the cue words multiple times. We increased
the number of cloud prompts from 64 to 1024, and fixed the number
of edge prompts at 512 tokens, each prompt was sampled from the ShareGPTv3
dataset.

As the Fig.\ref{fig:The-GPU-throughput}(a) shown, the PageAttention and RelayAttention implemented in the cloud side, with the system token increasing from 64 to 1024, the throughput was reduced by 50\%, for the PageAttention. The reason is that longer prompts hinder processing efficiency saturating the model's capacity for non-prompt content. This supports our conclusion that excessive prompt lengths degrade model performance. RelayAttention, however, maintains high throughput regardless of system token length due to an advanced pre-fill technique, which offsets load increases from longer prompts. EdgePrompt also employs a similar pre-fill technique to the payloads in the long system token case. Its peak throughput is comparable to that of the RelayAttention method, and as the number of filling rounds increases, RelayAttention spends more time in the process of initializing the parallel tensor, EdgePrompt gradually shows its advantage, as shown in Fig.\ref{fig:The-GPU-throughput}(a). 

As shown in Fig.\ref{fig:The-GPU-throughput}(b), when we deploy the Llama2-7b model on the edge device, EdgePrompt has a throughput advantage of about 10\% to 20\%. While this advantage extends to 50\% over PageAttention, EdgePrompt achieves much larger throughput, suggesting that EdgePrompt will more significantly outperform alternatives as the cloud-edge hardware performance gap widens.

\subsection{Interactive Service\label{subsec:Interactive-Serving}}

In the interactive service, the cloud side simulates the highly concurrent responses as chatbot. The metrics include response throughput ($req/s$) and latency ($s/Token$) for different request rates, quantities, and prompt lengths. In the throughput experiments, we fixed the number of requests sent as 1000 requests and set from 512 to 1024 cloud prompt tokens. Based on the poisson distribution involving the rate of request arrival, the experiment simulated the throughput and delay of the model under high loads and long cloud prompts.

\begin{figure}
\vspace{-0.2cm}
\includegraphics[scale=0.25]{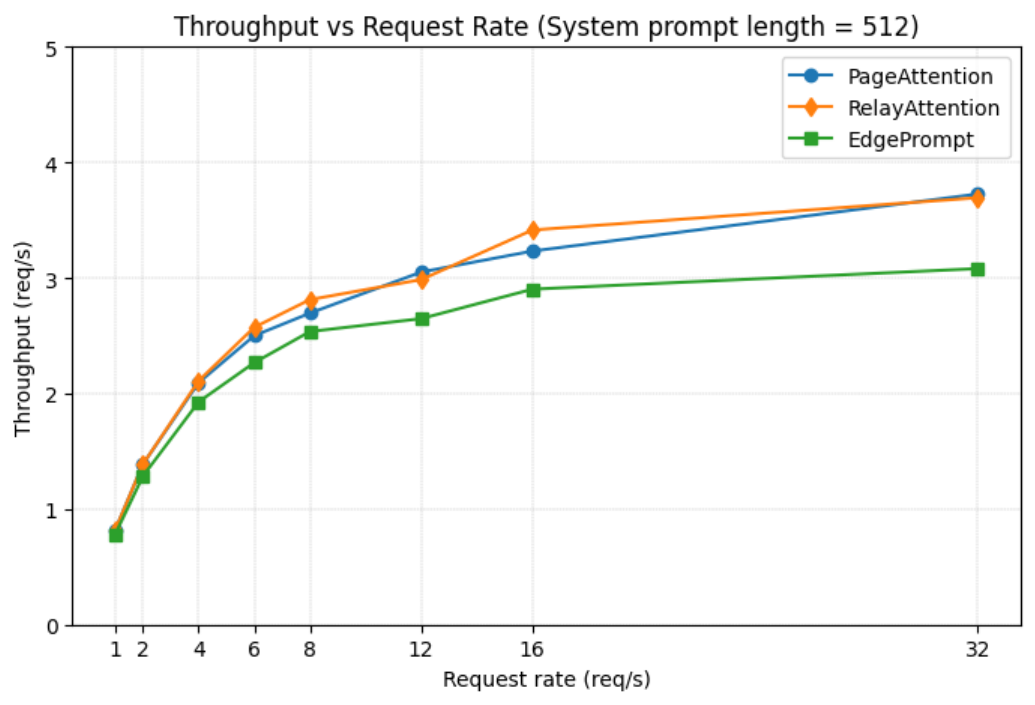}\includegraphics[scale=0.25]{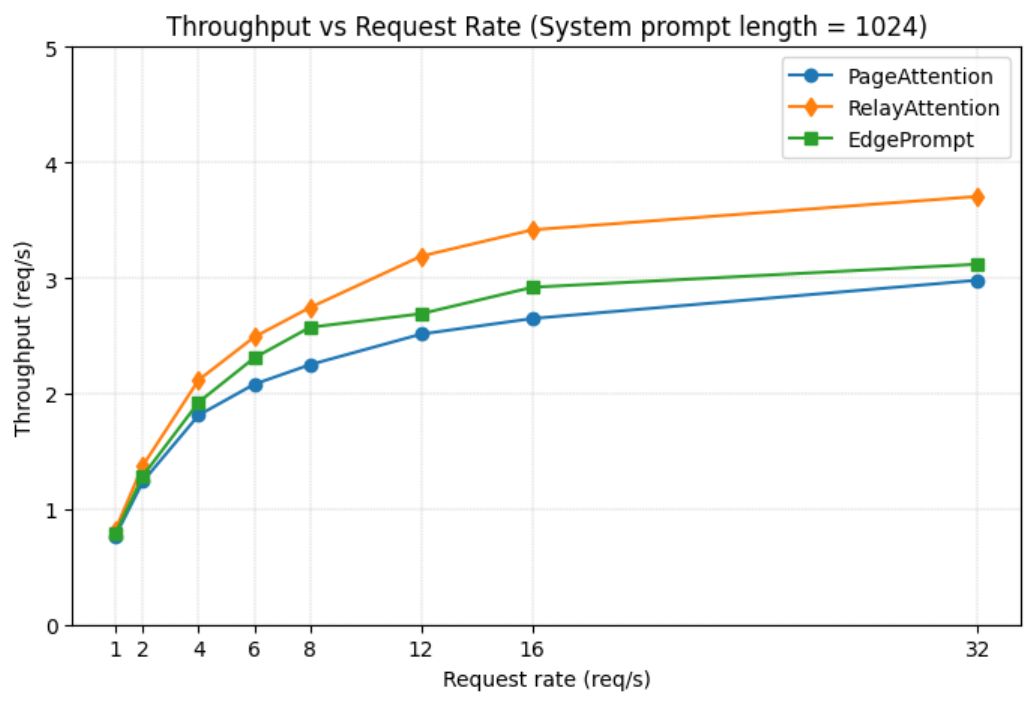}

\caption{Benchmark serving for throughput test under concurrent requests (Request
= 1000)}
\label{fig:Benchmark-serving-for-throughtput}

\vspace{-0.2cm}
\end{figure}

\begin{figure}
\includegraphics[scale=0.25]{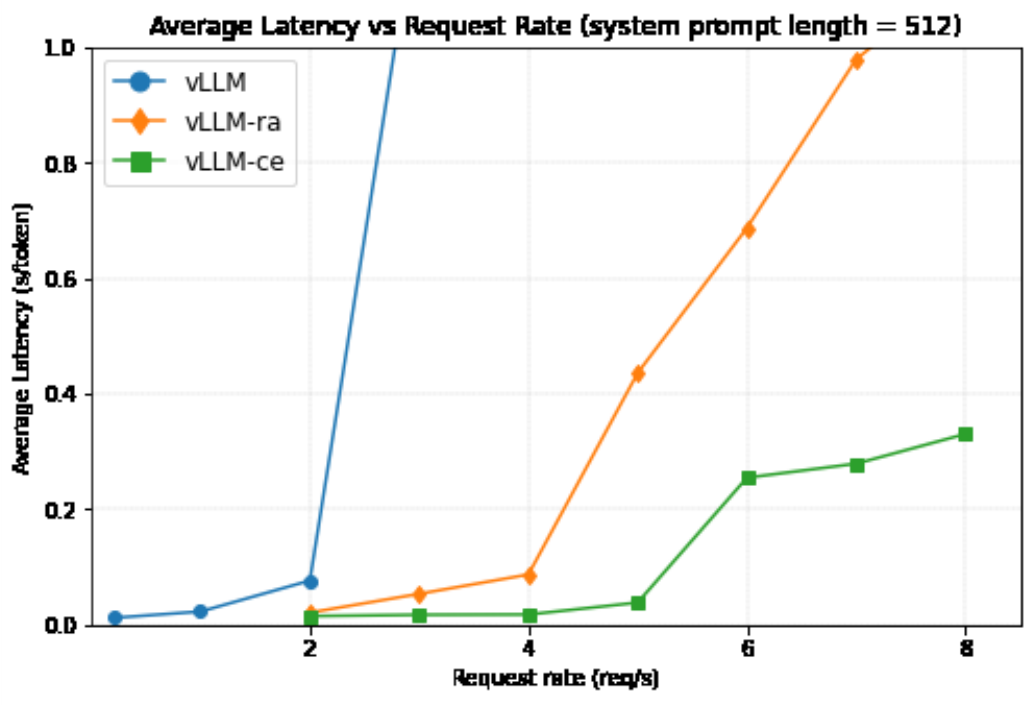}\includegraphics[scale=0.25]{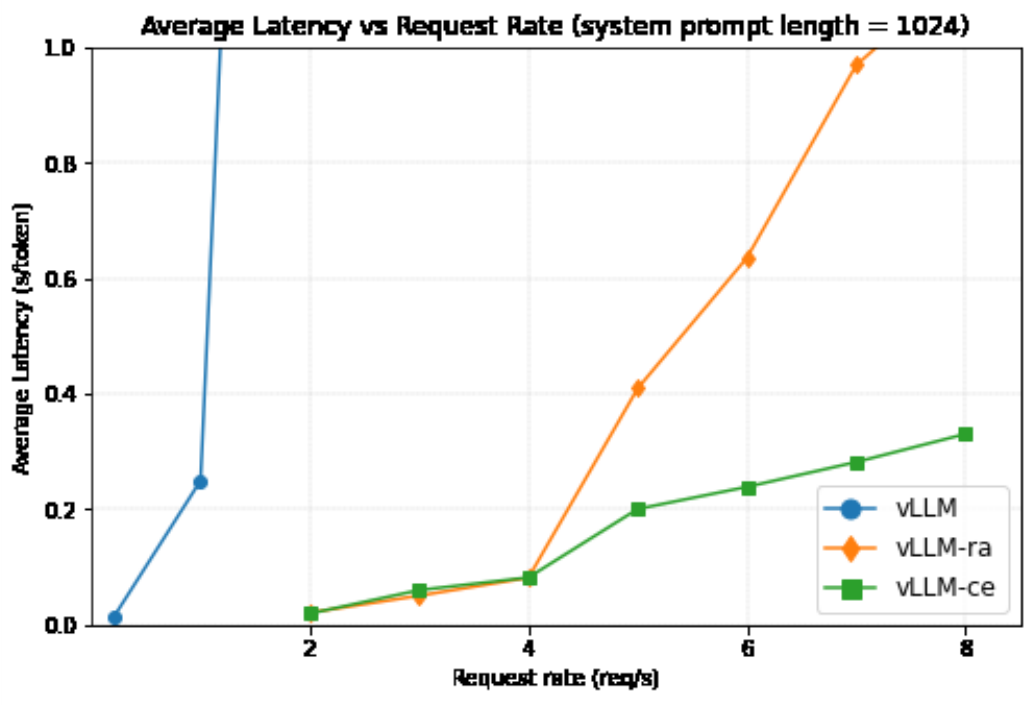}

\caption{Benchmark serving for latency test under concurrent request (Request
= 1000)}
\label{fig:Benchmark-serving-for-lantency-1024}

\vspace{-0.5cm}
\end{figure}

As shown in the Fig.\ref{fig:Benchmark-serving-for-throughtput}, with the request rate increases, the throughput of the EdgePrompt increases and exceeds that of the pageAttention method. This indicates that EdgePrompt has better throughput under high load and long system hints. Similarly, in Fig.\ref{fig:Benchmark-serving-for-lantency-1024}, we set the increasing rate and test the latency in the concurrent. We can see that the latency gradually increases as the number of concurrent requests increases. And EdgePrompt achieves better delay processing effect than RelayAttention in the continuously increasing concurrentrequests.

Combining the above two experiments, we can see that our EdgePrompt has better throughput and lower latency than the baseline method page attention in high concurrent request scenarios. Although compared with the RelayAttention method, our method does not achieve significant throughput advantage in the case of high concurrency, but with the improvement of concurrency speed, EdgePrompt method can still maintain low latency. 

\section{Conclusion\label{sec:Conclusion}}

This paper proposes a distributed privacy-enhanced LLMs inference scheme EdgePrompt based on cloud-edge collaborative architecture. The goal of this architecture is to optimize model throughput and
reduce request latency while ensuring robust privacy protection. A KV pair synchronization scheme is employed to sensitive data and synchronization efficiency. In addition, theoretical analysis and formula derivation provide a new basis for understanding the interactive dynamics and evolution of llm cloud edge collaborative reasoning.

Through extensive experimentation using varied GPUs and interaction methods, we evaluate performance of EdgePrompt across non-interactive throughput, high-concurrency scenarios, and privacy protection. The results demonstrate that EdgePrompt achieves efficient generation and effective load balancing while maintaining user privacy. Notably, EdgePrompt remains fully compatible with existing acceleration techniques, such as Quantification and parallel decoding, indicating that further performance enhancements are achievable with additional optimizations.

Despite our design, analysis, and optimization efforts, there is still room for improvement in this study. Our experiments focused on traditional throughput testing and did not fully explore the privacy and distributed interaction behavior of EdgePrompt, especially the transmission process and optimization problems in complex networks. Future research will focus on these aspects to refine EdgePrompt\textquoteright s applicability and resilience in distributed environments.

\vspace{12pt}

\bibliographystyle{unsrt}
\bibliography{Edge_collebrated_inference}

\begin{thebibliography}{10}

\bibitem{1}
Yuanchun Li, Hao Wen, Weijun Wang, Xiangyu Li, Yizhen Yuan, Guohong Liu, Jiacheng Liu, Wenxing Xu, Xiang Wang, Yi~Sun, et~al.
\newblock Personal \textsc{LLM} agents: Insights and survey about the capability, efficiency and security.
\newblock {\em arXiv preprint arXiv:2401.05459}, 2024.

\bibitem{28}
Aakanksha Chowdhery, Sharan Narang, Jacob Devlin, Maarten Bosma, Gaurav Mishra, Adam Roberts, Paul Barham, Hyung~Won Chung, Charles Sutton, Sebastian Gehrmann, et~al.
\newblock Palm: Scaling language modeling with pathways.
\newblock {\em Journal of Machine Learning Research}, 24(240):1--113, 2023.

\bibitem{58_network-LLM}
Yudong Huang, Hongyang Du, Xinyuan Zhang, Dusit Niyato, Jiawen Kang, Zehui Xiong, Shuo Wang, and Tao Huang.
\newblock Large language models for networking: Applications, enabling techniques, and challenges.
\newblock {\em IEEE Network}, pages 1--1, 2024.

\bibitem{42_leveraging_ai_review}
Tom Goethals, Bruno Volckaert, and Filip De~Turck.
\newblock Enabling and leveraging \textsc{AI} in the intelligent edge: A review of current trends and future directions.
\newblock {\em IEEE Open Journal of the Communications Society}, 2:2311--2341, 2021.

\bibitem{41}
Bomin Mao, Jiajia Liu, Yingying Wu, and Nei Kato.
\newblock Security and privacy on \textsc{6G} network edge: A survey.
\newblock {\em IEEE communications surveys \& tutorials}, 25(2):1095--1127, 2023.

\bibitem{12-llmservey}
Zhihang Yuan, Yuzhang Shang, Yang Zhou, Zhen Dong, Chenhao Xue, Bingzhe Wu, Zhikai Li, Qingyi Gu, Yong~Jae Lee, Yan Yan, et~al.
\newblock \textsc{LLM} inference unveiled: Survey and roofline model insights.
\newblock {\em arXiv preprint arXiv:2402.16363}, 2024.

\bibitem{llm_qat}
Zechun Liu, Barlas Oguz, Changsheng Zhao, Ernie Chang, Pierre Stock, Yashar Mehdad, Yangyang Shi, Raghuraman Krishnamoorthi, and Vikas Chandra.
\newblock \textsc{LLM-QAT}: Data-free quantization aware training for large language models.
\newblock {\em arXiv preprint arXiv:2305.17888}, 2023.

\bibitem{lut_gemm}
Gunho Park, Baeseong Park, Minsub Kim, Sungjae Lee, Jeonghoon Kim, Beomseok Kwon, Se~Jung Kwon, Byeongwook Kim, Youngjoo Lee, and Dongsoo Lee.
\newblock \textsc{LUT-GEMM}: Quantized matrix multiplication based on \textsc{LUT}s for efficient inference in large-scale generative language models.
\newblock {\em arXiv preprint arXiv:2206.09557}, 2022.

\bibitem{63_roofline_model}
Samuel Williams, Andrew Waterman, and David Patterson.
\newblock Roofline: an insightful visual performance model for multicore architectures.
\newblock {\em Commun. ACM}, 52(4):65--76, April 2009.

\bibitem{5-SS}
Charlie Chen, Sebastian Borgeaud, Geoffrey Irving, Jean-Baptiste Lespiau, Laurent Sifre, and John Jumper.
\newblock Accelerating large language model decoding with speculative sampling.
\newblock {\em arXiv preprint arXiv:2302.01318}, 2023.

\bibitem{65-edge}
Wentao Zhao, Wenpeng Jing, Zhaoming Lu, and Xiangming Wen.
\newblock Edge and terminal cooperation enabled \textsc{LLM} deployment optimization in wireless network.
\newblock In {\em 2024 IEEE/CIC International Conference on Communications in China (ICCC Workshops)}, pages 220--225, 2024.

\bibitem{66_hybrid}
Zixu Hao, Huiqiang Jiang, Shiqi Jiang, Ju~Ren, and Ting Cao.
\newblock Hybrid \textsc{SLM} and \textsc{LLM} for edge-cloud collaborative inference.
\newblock In {\em Proceedings of the Workshop on Edge and Mobile Foundation Models}, EdgeFM '24, pages 36--41, New York, NY, USA, 2024. Association for Computing Machinery.

\bibitem{67_EdgeLLM}
Daliang Xu, Wangsong Yin, Hao Zhang, Xin Jin, Ying Zhang, Shiyun Wei, Mengwei Xu, and Xuanzhe Liu.
\newblock Edge\textsc{LLM}: Fast on-device \textsc{LLM} inference with speculative decoding.
\newblock {\em IEEE Transactions on Mobile Computing}, pages 1--18, 2024.

\bibitem{6}
Woosuk Kwon, Zhuohan Li, Siyuan Zhuang, Ying Sheng, Lianmin Zheng, Cody~Hao Yu, Joseph Gonzalez, Hao Zhang, and Ion Stoica.
\newblock Efficient memory management for large language model serving with \textsc{P}aged\textsc{A}ttention.
\newblock In {\em Proceedings of the 29th Symposium on Operating Systems Principles}, pages 611--626, 2023.

\bibitem{7-MLsys}
In~Gim, Guojun Chen, Seung-seob Lee, Nikhil Sarda, Anurag Khandelwal, and Lin Zhong.
\newblock Prompt cache: Modular attention reuse for low-latency inference.
\newblock In P.~Gibbons, G.~Pekhimenko, and C.~De Sa, editors, {\em Proceedings of Machine Learning and Systems}, volume~6, pages 325--338, 2024.

\end{thebibliography}

\end{document}